\documentstyle[preprint,aps]{revtex}
\begin{document}
\draft
\title{Measurement of the Reaction $^{12}{\rm C}(\nu_\mu,\mu^-) {\rm X}$
Near Threshold}
\author{M. Albert$^{12}$, C. Athanassopoulos$^{13}$, L. B. Auerbach$^{13}$,
 D. Bauer$^3$,
R. Bolton$^7$, B. Boyd$^9$, R. L. Burman$^7$,
I. Cohen$^6$,\\D. O. Caldwell$^3$, B. D. Dieterle$^{10}$, J. Donahue$^7$,
A. M. Eisner$^4$, A. Fazely$^{11}$,
F. J. Federspiel$^7$, G. T. Garvey$^7$,\\R. M. Gunasingha$^8$,
V. Highland$^{13}$, J. Hill$^{12}$, R. Imlay$^8$, K. Johnston$^{9}$,
W. C. Louis$^7$,
A. Lu$^3$, A. K. Mann$^{12}$,\\
J. Margulies$^{13}$,  K. McIlhany$^{1}$, W. Metcalf$^8$,
R. A. Reeder$^{10}$, V. Sandberg$^7$,
M. Schillaci$^7$, D. Smith$^{5}$,\\I. Stancu$^{1}$, W. Strossman$^{1}$,
 M. K. Sullivan$^4, $G. J. VanDalen$^{1}$,
W. Vernon$^{2,4}$, Y-X. Wang$^4$, D. H. White$^7$, \\D. Whitehouse$^7$,
D. Works$^{13}$, Y. Xiao$^{13}$,
S. Yellin$^3$}
\address{$^{1}$University of California, Riverside, CA 92521}
\address{$^{2}$University of California, San Diego, CA 92093}
\address{$^3$University of California, Santa Barbara, CA 93106}
\address{$^4$University of California
Intercampus Institute for Research at Particle Accelerators,
Stanford, CA 94309}
\address{$^{5}$Embry Riddle Aeronautical University, Prescott, AZ 86301}
\address{$^6$Linfield College, McMinnville, OR 97128}
\address{$^7$Los Alamos National Laboratory, Los Alamos, NM 87545}
\address{$^8$Louisiana State University, Baton Rouge, LA 70803}
\address{$^{9}$Louisiana Tech University, Ruston, LA 71272}
\address{$^{10}$University of New Mexico, Albuquerque, NM 87131}
\address{$^{11}$Southern University, Baton Rouge, LA 70813}
\address{$^{12}$University of Pennsylvania, Philadelphia, PA 19104}
\address{$^{13}$Temple University, Philadelphia, PA 19122}

\date{\today}
\maketitle
\begin{abstract}
The reaction
$^{12}{\rm C}(\nu_\mu,\mu^-) {\rm X}$ has
been measured near threshold
using a $\pi ^+$ decay-in-flight $\nu_\mu$ beam from the Los Alamos
Meson Physics Facility and a massive liquid
scintillator neutrino detector (LSND).
In
the energy region $123.7 < {\rm E}_\nu < 280$ MeV,
the measured spectral shape is consistent with that expected
from the Fermi Gas Model.
However, the measured flux--averaged inclusive cross section
(~$(8.3 \pm 0.7 {\rm stat.} \pm 1.6 {\rm syst.})
\times 10^{-40} {\rm cm}^2$~) is more than a
factor of 2 lower than that
predicted by the Fermi Gas Model and by a
recent random phase approximation calculation.
\end{abstract}

There has been little information to date on low energy
neutrino--nucleus scattering despite its potential
application to nuclear structure studies.  This process
principally involves axial--vector nuclear currents and
consequently provides different information than low energy
electron-nucleus scattering, which is sensitive only to
polar--vector currents.  While the coupling of the W$^{\pm}$
to a free nucleon is well understood at low Q$^2$
(Q$^2  < 1$ GeV$^2$),
calculation of the $(\nu,{\em l}^{\pm})$ inclusive cross section
from a nucleus is beyond the capabilities of present models.
The Fermi Gas Model\cite{Fermi_gas}, for example,
works well at higher Q$^2$ but
is not expected to accurately reflect the behavior of nuclei
probed at low momentum transfer.
There are a variety of important strong--interaction dynamical effects
that are not readily incorporated into models sufficiently global to
reproduce inclusive cross sections.
These inclusive charged--current processes,
${\rm A}(\nu,{\em l}^{\pm}) {\rm X}$,
have taken on new
importance in recent years as
as they are central to the
detection process
in several active neutrino detectors.

We report here
measurements of the salient features of the reaction
$^{12}{\rm C}(\nu_\mu,\mu^-){\rm X}$ from
threshold (123.7 MeV) to 280 MeV neutrino energy.
The data were obtained in the
initial run of the Liquid Scintillator Neutrino Detector (LSND) at
the Los Alamos Meson Physics Facility (LAMPF).

LSND is a cylindrical imaging {$\check{\rm C}$erenkov} detector
9~m long and 6~m in diameter with
its axis horizontal.
 It consists of
 197~m$^3$ of mineral oil viewed by
1220 20 cm$\phi$ photomultiplier tubes (PMTs), which cover 24.8\%
of the detector's inner surface.  A small amount of scintillator
(0.031~g/l~butyl-PBD) is dissolved in the mineral oil,
so that both the scintillation and
{$\check{\rm C}$erenkov} light produced by charged particles may be
observed and resolved\cite{Reeder}.
For a given
amount of detected light, the ratio of light in the
{$\check{\rm C}$erenkov} cone
to isotropic light (which includes wave--shifted
{$\check{\rm C}$erenkov} light)
facilitates identification of
particle type.  For highly relativistic
particles, this ratio is approximately 1:4.
The spatial origin of the light associated with an event
can be localized to within 25 cm rms using
the photon arrival times
at each hit PMT. For electrons, the relationship
between the total detected PMT charge and
particle energy is determined by the 52.8 MeV
endpoint of electrons from the decay of stopping cosmic
ray muons (the Michel spectrum).
In this
spectrum, 32 PEs per MeV are detected,
where a PE is defined as the peak of the PMT response to a
single photoelectron.  The energy resolution
at the endpoint of the spectrum is 6\%.
The corresponding relationship between charge and energy
for
other particle types is determined
by a GEANT-based Monte Carlo simulation
of LSND\cite{MC} which reproduces the observed
Michel spectrum and incorporates
data from an exposure
of a sample of LSND scintillator to protons and electrons of known
energy\cite{Reeder}.

The midpoint of the LSND detector is located
29~m downstream of the LAMPF A--6
proton beam dump at $12^\circ$ from
the axis of the proton beam.
LSND is surrounded (except for the bottom) by a highly efficient
cosmic ray veto counter\cite{VETO}, which is crucial for eliminating
backgrounds
that would otherwise arise from the 4kHz rate of cosmic muons in the
detector.

The trigger requires signals above threshold in at least 100 of the
detector PMTs, and fewer than 6 hit tubes in the  veto counter.
 When this trigger is satisfied,
the event is read out along with every other event that fired either
the veto counter or at least 18 PMTs in the detector within the previous
50 $\mu$s.
To remove the burden on
the data acquisition system of recording decay electrons
from stopping cosmic muons, the trigger is disabled for
7 muon lifetimes following each firing of the veto.

 For the data reported here,
 a 780~MeV proton beam at 600--700 $\mu$A was delivered at a
7.1\% duty factor
to the A--6 proton beam dump.  The integrated intensity was 1625 C.
The beam dump configuration
consisted of a 20~cm long water target, several inserts used for isotope
production, and a copper proton beam stop.
The water target serves as the main source of pions for both the
decay--in--flight (DIF) and decay--at--rest (DAR)
neutrino beams, with a smaller
contribution to the DAR neutrinos
arising from the beam stop directly.  Because
of the 123.7 MeV threshold for the
$^{12}{\rm C}(\nu_\mu,\mu^-) {\rm X}$ reaction, only DIF neutrinos
contribute.

The DIF neutrino flux is calculated by a
Monte Carlo simulation of the beam dump\cite{beam_mc}, and
includes the flux from the two thin targets well upstream of the beam
dump, whose contributions are significant only at the
highest neutrino energies. Because the decay chain
$
\pi^+ \rightarrow \mu^+ + \nu_\mu$
followed by
$\mu^+
 \rightarrow  {\rm e}^+ +
\nu_{\rm e} + {\overline \nu_{\mu}}
$
is dominant, the integrated
neutrino flux from $\pi^+$ DIF may be constrained
by measurements of the neutrino flux from $\mu^+$ DAR. The
Monte--Carlo--calculated flux from DAR has been verified in an
independent measurement \cite{beam_mock_up} to an accuracy
of $\pm 8\%$
and confirmed in an experiment that
measured $\nu_{\rm e}$ elastic scattering from electrons\cite{e225}
to an accuracy of $\pm 15\%$(stat.) $\pm 9\%$(syst.).
We estimate a systematic uncertainty of $\pm 15\%$ in the
DIF neutrino flux over the relevant energy
interval of 123.7 to 280~MeV.  This flux distribution is shown in Fig.~1.

The
quasielastic process $^{12}{\rm C}(\nu_\mu,\mu^-){\rm X}$
produces a muon in the interior of the detector, usually accompanied
by a promptly ejected proton.  The muon + proton signal is
followed by an electron from the
($\tau = 2.03{\rm \mu s}$) decay of that muon in mineral
oil \cite{muon_capture}.
Thus we select pairs of events occurring within 17$\mu$s of
each other
 and reconstructing within 200~cm  of each other.
Requiring less than 4 hit tubes in the veto counter
suppresses the cosmic ray muon contribution by a factor of
$5\times10^{-5}$.
 The majority of the remaining cosmic--ray--induced
events
are eliminated by imposing the following additional
criteria:
First, the number of PEs detected for the first event
(the $\mu^- + p$ candidate) is required
to be less than the maximum expected from a
${\rm C}(\nu_\mu,\mu^-)X$ event, given the flux shown in Fig.~1.
Second, the energy associated with
the electron candidate is required to be less than
60 MeV, and above the endpoint energy (13.6 MeV) of $^{12}{\rm B}$
beta decay.
($^{12}{\rm B}$
is formed in the detector
by the capture of cosmic ray $\mu^-$  on $^{12}{\rm C}$.  This
cut is accomplished by requiring the electron to fire more than 250
PMTs.)
Finally,
both events are required to have reconstructed positions
within the central 108~m$^3$ fiducial volume.
The efficiency of these selection criteria is $34\pm 4$\%
(see Table \ref{Eff}).
Because of greater background for the lowest energy muons,
 tighter
selection criteria (listed in Table \ref{Eff}) were applied to
these events, reducing the efficiency for those
muons to $25\pm 3$\%.
(Less than 10\% of the signal is in this lowest energy region.)
A total of 270~events pass these selection criteria.

In this sample, the most important beam--related background
is from $\pi^-$ DIF followed by
$\overline{\nu_\mu} + p \rightarrow \mu^+ + {\rm n}$. This process was
calculated (using the cross section and form factors
in ref. \cite{LS})
to give $14\pm5$ events.  Contributions from
the two other neutrino--related backgrounds,
$\overline{\nu_\mu} + ^{12}{\rm C} \rightarrow \mu^+ + {\rm X}$, and
$\nu_\mu +  ^{13}{\rm C} \rightarrow \mu^- + {\rm X}$
are closely related to the cross section being measured,
and were estimated to be less than 4\% of the observed yield.
{}From the number of events recorded with the beam off, we
infer that
$40\pm 2$ of the 270 events are
due to the cosmic--ray--induced background that passes the
selection criteria.
All histograms shown for this sample have this beam--off
contribution subtracted on a bin--by--bin basis.

Fig. 2a shows the spectrum of time differences between the muon
and electron
candidates in our final sample.  This figure implies a mean muon
lifetime
of $2.1\pm 0.2 \mu {\rm s}$,  consistent with expectation.
The spatial separation
between the $\mu$ and e candidates in each pair is shown in
Fig.~2b. The muons are distributed uniformly
throughout the fiducial volume.
The energy spectrum
of the decay electrons is shown
in Fig.~2c.
To demonstrate that this energy
spectrum is representative of that produced
by the decay of muons in LSND, the (normalized) energy spectrum
of electrons from the decay of stopped cosmic muons is also shown.

The charge measured in LSND from  ${\rm C}(\nu_\mu,\mu^-){\rm X}$
events
comes from both the $\mu^-$ and the (usually) ejected proton.
The light output as a function of particle energy differs
for the semi-relativistic muons and
non-relativistic protons. A 180~MeV incident neutrino, for example,
produces a quasielastic event with total PMT charge between 400~and
600~PEs, depending on the sharing of available kinetic energy between
the final state $\mu^-$ and $p$\cite{Reeder}.
Events with summed PMT~charge
greater than 1500~PEs correspond to neutrino energies above 230~MeV.
Because it is not possible to accurately recover the
energy of the muon or the incident neutrino given only
the total charge detected, we show in Fig.~3
the spectrum of collected charge for the quasielastic events,
measured in terms of PEs.
The collected charge spectrum predicted by a
Coloumb-corrected Fermi Gas Model (FGM) \cite{Fermi_gas},
normalized to the total number of observed  events,
is superimposed on the data in Fig.~3.  For additional
comparison,
the calculated\cite{LS} charge spectrum of $\nu_\mu$ on free neutrons,
also normalized to the data, is shown in the same figure.
(These calculated spectra have a systematic uncertainty in their
charge
scales, arising mostly from uncertainty in the
amount of light produced by highly--ionizing
low--energy protons.  The scale uncertainty for 180 MeV neutrinos
is $\pm 10\%$; the effect of any such rescaling factor increases at
lower energies.)
The shapes of both calculated spectra
agree with the  shape of the
experimental data.
A similar level of agreement was also obtained
with a low statistics sample
of quasielastic events reported by the E645 collaboration
at LAMPF \cite{shape}.
The general agreement between the data
and
both the free neutron and FGM calculations in Fig.~3 indicates
that the spectrum shape is not particularly sensitive to the nuclear
dynamics which these models do not include.

The yield for the exclusive reaction to the ground state,
$^{12}{\rm C}(\nu_\mu,\mu^-)^{12}{\rm N}$(g.s.),
is well predicted, as it depends only on form factors
measured in beta decay, muon capture, and electron
scattering.  Fortuitously, this yield can be measured because the
 $^{12}$N ground state is the only  $^{12}$N level stable against
strong decays, so its subsequent beta decay
(${\rm E_o} = 16.3$ MeV, $\tau = 15.9$ms)
serves to uniquely identify it.
We presently observe six events for this reaction and the subsequent beta
decay, while seven are expected from
reliable calculations of this ground state transition \cite{Kolbe}.
This gives us some confidence that our calculation of fluxes and
detection efficiencies are correct.

The net number of {\em inclusive }
$^{12}{\rm C}(\nu_\mu,\mu^-){\rm X}$
events detected (after subtracting the beam--off and
three beam--related backgrounds)
is $210 \pm 17$ events.
This  corresponds to
a flux--averaged inclusive cross section of
 $(8.3 \pm 0.7 {\rm stat.} \pm 1.6 {\rm syst.})
\times 10^{-40} {\rm cm}^2$ in
the energy region $123.7 < {\rm E}_\nu < 280$ MeV.
The flux--weighted average neutrino energy is
$<{\rm E}_\nu> = 180$ MeV.
This average cross section is lower than that obtained
using the FGM
($24\times 10^{-40} {\rm cm}^2$)
and a recent continuum random phase approximation
(RPA) calculation\cite{Kolbe}
($20 \times 10^{-40} {\rm cm}^2$) evaluated with the flux shown in Fig 1.
An earlier calculation \cite{Kim} using the measured $\mu^-$
capture rates on
$^{12}$C along with a closure approximation gives a flux--averaged
cross section of $11 \times 10^{-40} {\rm cm}^2$, in
agreement with our measurement; however, it is not clear
if important contributions from partial waves with $l\ge2$ were properly
accounted for.
Our measurement is substantially lower than
the average cross section reported by
an earlier experiment\cite{Koetke} involving a brief exposure of
a less massive,
segmented detector to a different (and slightly higher energy)
neutrino beam at LAMPF.  This previous measurement reported a
visible energy spectrum significantly softer than that predicted
by the FGM.

One may also compare the measurements of
$^{12}{\rm C}(\nu_{\rm e},{\rm e}^-){\rm X}$ (made using
$\nu_{\rm e}$ from muon DAR) to model predictions.  Two
measurements\cite{e645} \cite{KARMEN} of the exclusive
cross section to the ground state of
$^{12}{\rm N}$ are in good agreement
with one another and with various calculations
of the expected yield \cite{Kolbe} \cite{Kim} \cite{Donn}.
Agreement between measured
and predicted cross sections to excited states of
$^{12}{\rm N}$
is less well established.  The measured values
($3.6\pm 2.7 \times 10^{-42} {\rm cm}^2$\cite{e645}, and
 $6.4\pm 2.0 \times 10^{-42} {\rm cm}^2$\cite{KARMEN})
span a factor of
two, but agree within quoted errors.  Predictions for
this yield ( $6.3 \times 10^{-42}$ (RPA) , and
$3.7 \times 10^{-42}$\cite{Donn}) span the same factor
of two.

Thus at present there is agreement to $\approx$ a factor two
among simple FGM, continuum RPA calculations, and the inclusive
cross sections observed in low energy neutrino reactions.
However,
the measurement reported here is significantly lower than model
predictions, indicating the presence of nuclear effects
important at low energy
that are not accounted for in these models.

\paragraph*{Acknowledgements}

The authors gratefully acknowledge the technical contributions of
 V.~Armijo, K.~Arndt, D.~Callahan,
B.~Daniel, S.~Delay, C.~Espinoza, C.~Gregory, D.~Hale, G.~Hart,
W.~Marterer, and
T.~N.~Thompson
 to the construction
and operation of LSND. We also want to thank
 the students
G.~Anderson, C.~Ausbrooks, B.~Bisbee, L.~Christofek, D.~Evans,
J.~George, B.~Homann,
R.~Knoesel, S.~Mau, T.~Phan, F.~Schaefer,
M.~Strickland, M.~Volta,
S.~Weaver, and K.~Yaman for their help in making the
detector operational.

This work was supported by the U. S. Department of Energy and by
the National Science Foundation.


%
%

\begin{figure}
\caption{The calculated energy spectrum of muon neutrinos from the
decay--in--flight beam.}
\end{figure}

\begin{figure}
\caption{(a)Time difference between muon and electron candidates.
An exponential fit yields a lifetime of
$2.1\pm0.2 \mu {\rm s}$.  Muons that live less than
$0.7 \mu {\rm s}$ do not appear in the sample because of the
time required for the trigger to reset.
(b) Distance between the reconstructed positions of the muon and electron
in quasielastic events.  (c) Energy of the electron from muon decay.
Data points show the electron energy spectrum from the
decay of quasielastically produced muons.  The histogram shows
the spectrum obtained from a sample of stopping cosmic muons.
Data points have beam-off contributions subtracted bin--by--bin
in all three plots.}
\end{figure}

\begin{figure}
\caption{Data points with error bars show the detected charge distribution
of quasielastic events ($\overline{\nu_\mu}  {\rm p}$ and beam--off
contributions subtracted bin--by--bin)
compared with that predicted by the Fermi Gas Model[1]
scaled to the
data (solid line), and the predicted spectrum from scattering on free
neutrons[9],
scaled to the data (dashed line).
The lowest energy events correspond to neutrinos at the threshold for
${\rm C}(\nu_\mu,\mu^-){\rm X}$; the highest to neutrinos of 250 MeV
and greater.}
\end{figure}

%

\begin{table}
\caption{The efficiencies for selection of quasielastic events. As seen
in Figure 2,
the efficiencies for the spatial and temporal coincidences are
essentially 100\%.}
\label{Eff}
\begin{tabular}{lr}
Source&Efficiency\\
\tableline
veto counter inactive&$0.77\pm0.02$\\
computer ready&$0.97\pm0.01$\\
$\mu^-$ not captured&$0.92\pm0.01$\\
$\mu^-$ lives longer than 0.7~$\mu {\rm s}$ &$0.71\pm0.04$\\
$\mu^-$ and e$^-$ in fiducial volume&$0.78\pm0.08$\\
e$^-$ fires more than 250 PMTs&$0.90\pm0.02$\\
\tableline
Overall Efficiency for E$_{vis} \ge 140$ PE &$0.34\pm0.04$\\
\tableline
Additional cuts for E$_{vis} < 140$ PE:&\\
no cosmic muon in 51 $\mu$s prior to e$^-$&$0.88\pm0.01$\\
particle identification on electron
\tablenote{Efficiency for electron identification was determined
using the electrons from the decay of muons produced
in the higher energy quasielastic events.}
&$0.86\pm0.03$\\
\tableline
Overall Efficiency for E$_{vis} < 140$ PE &$0.25\pm0.03$\\
\end{tabular}
\end{table}

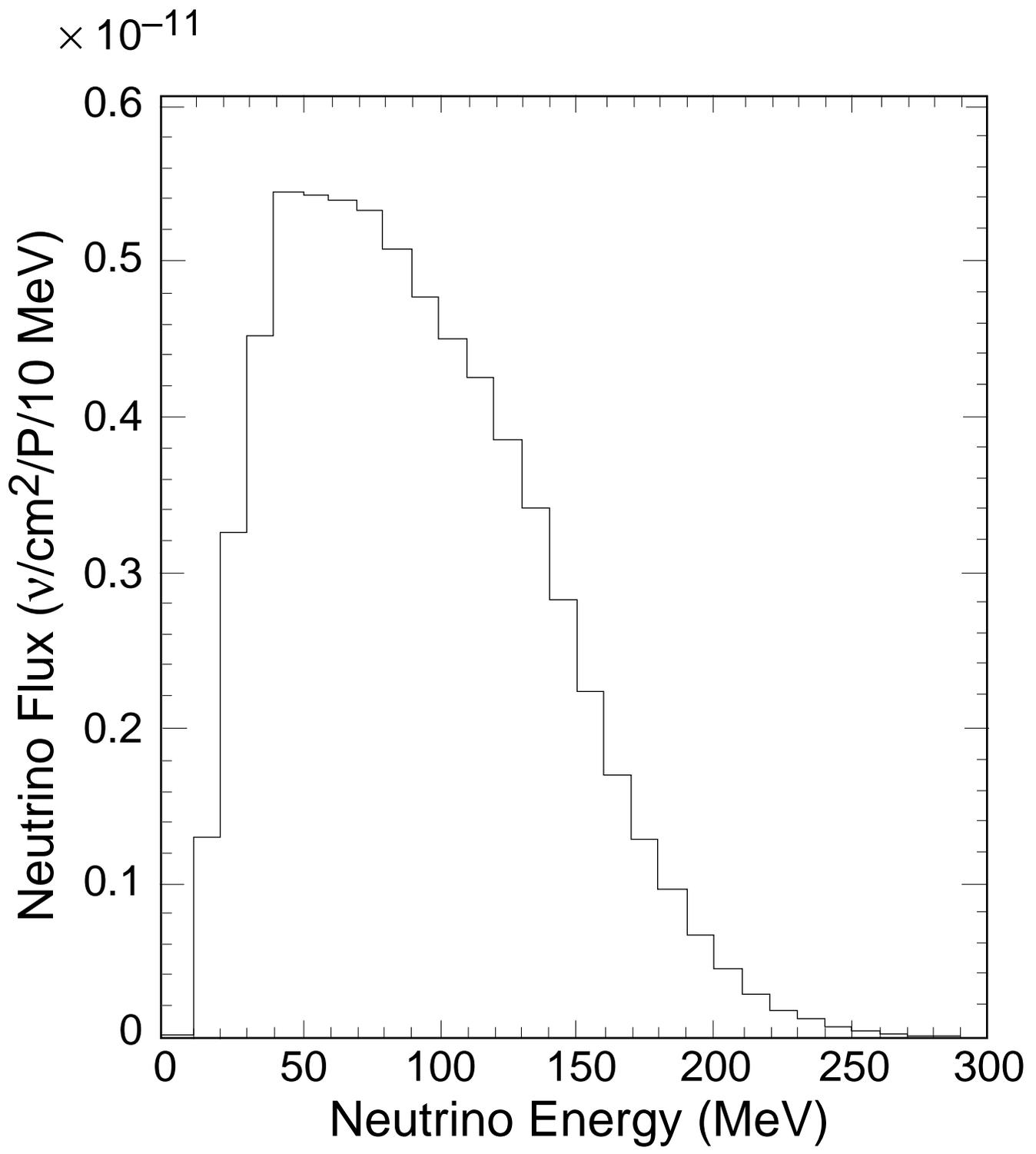
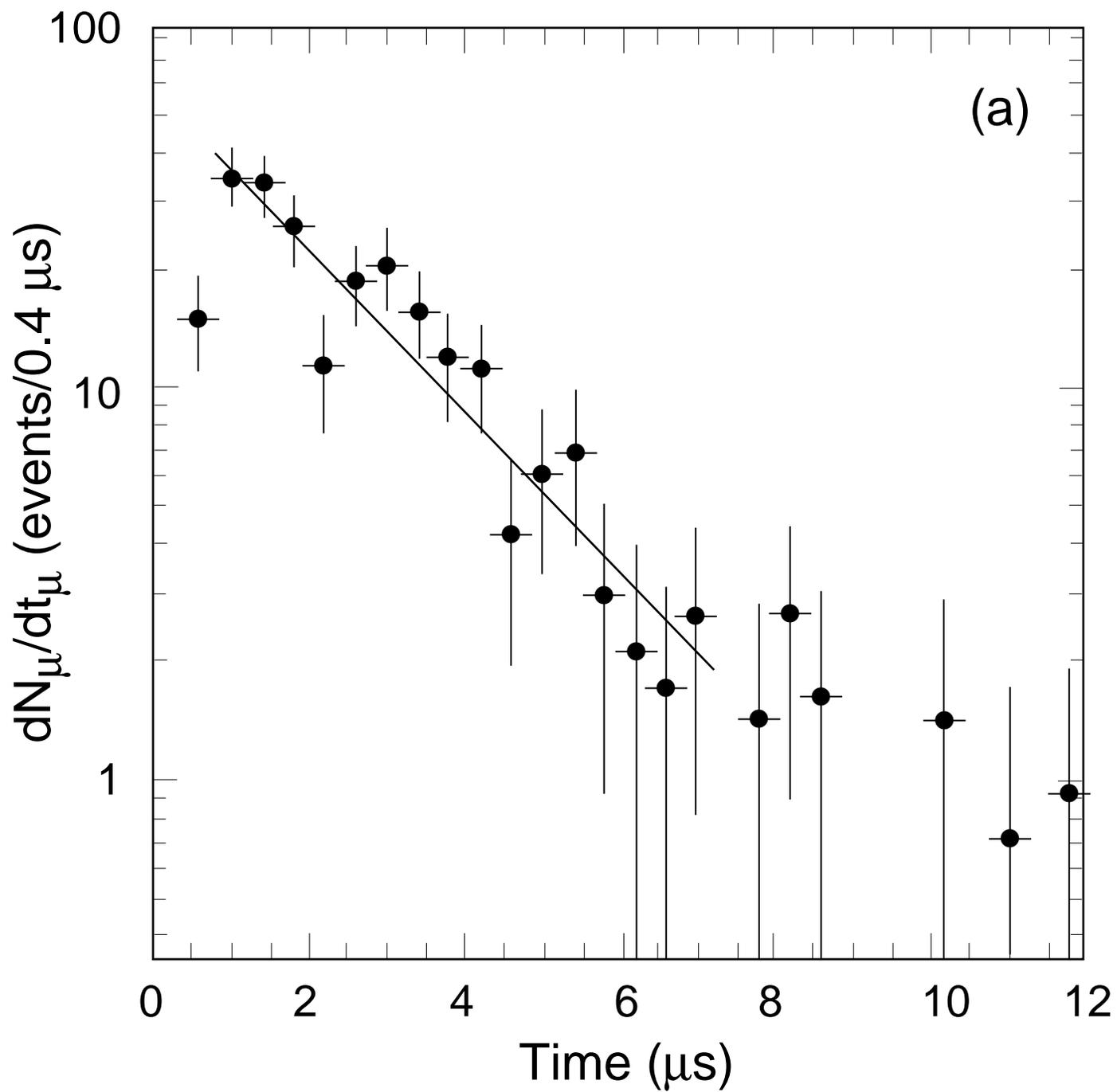
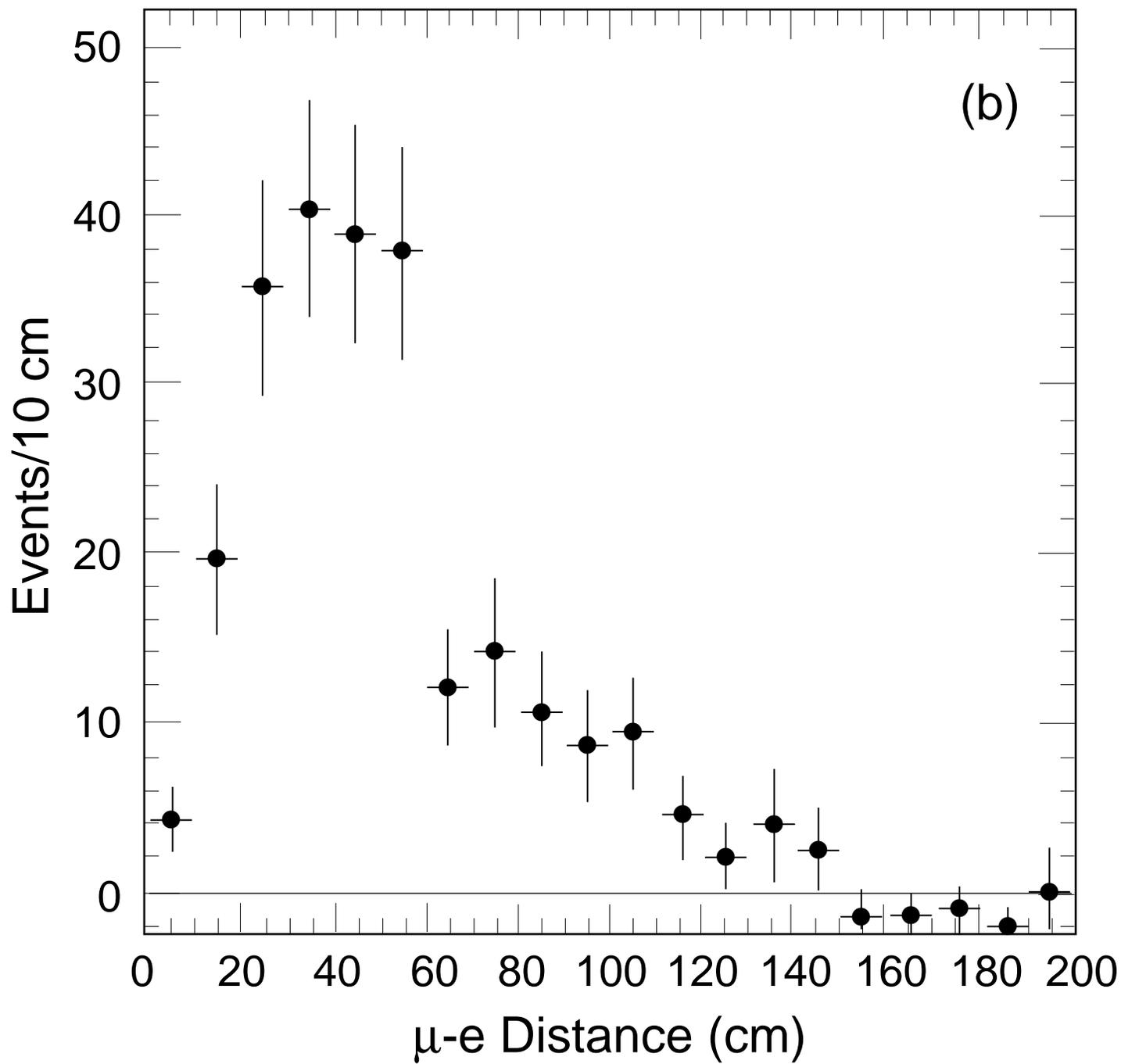
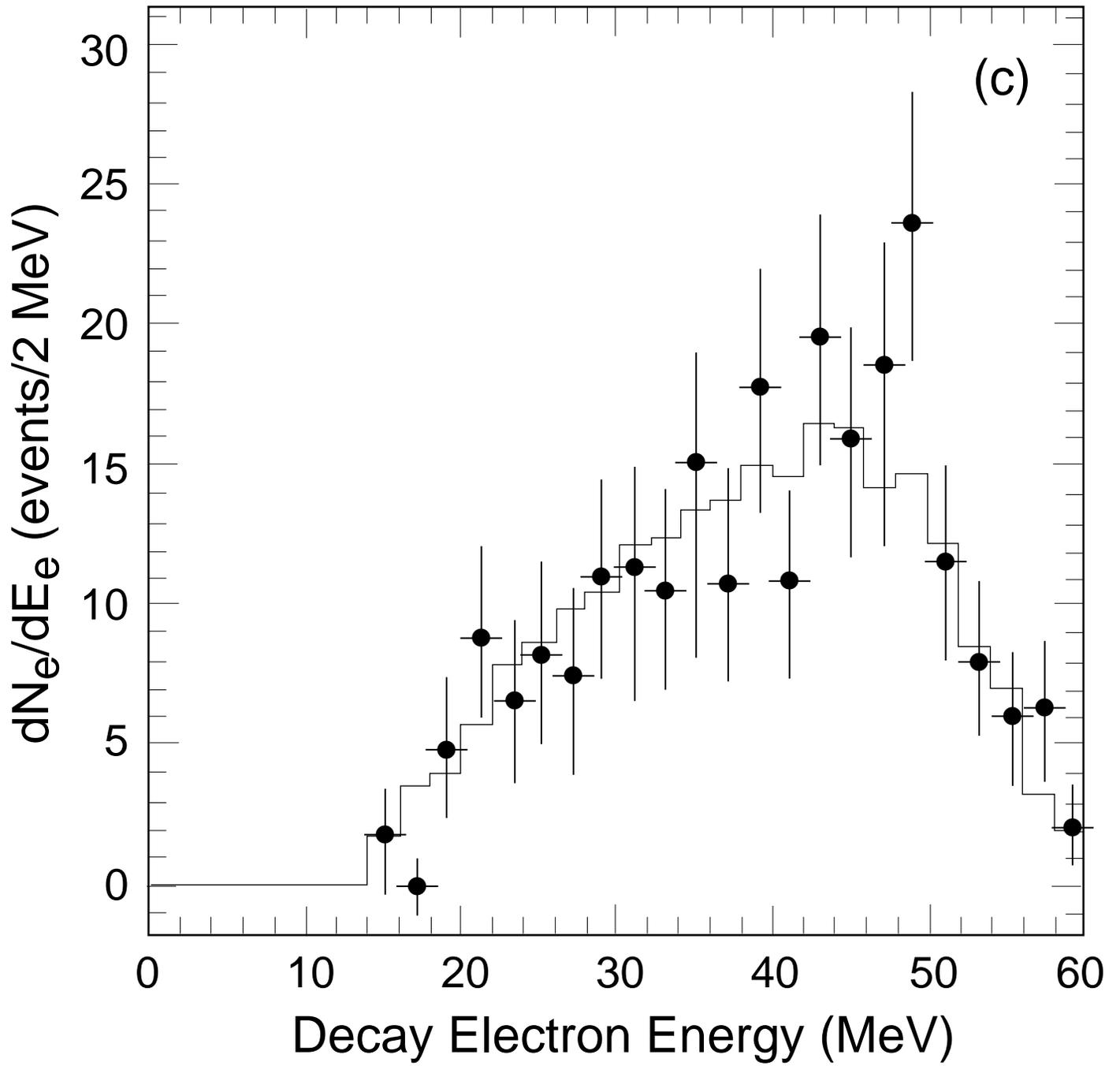
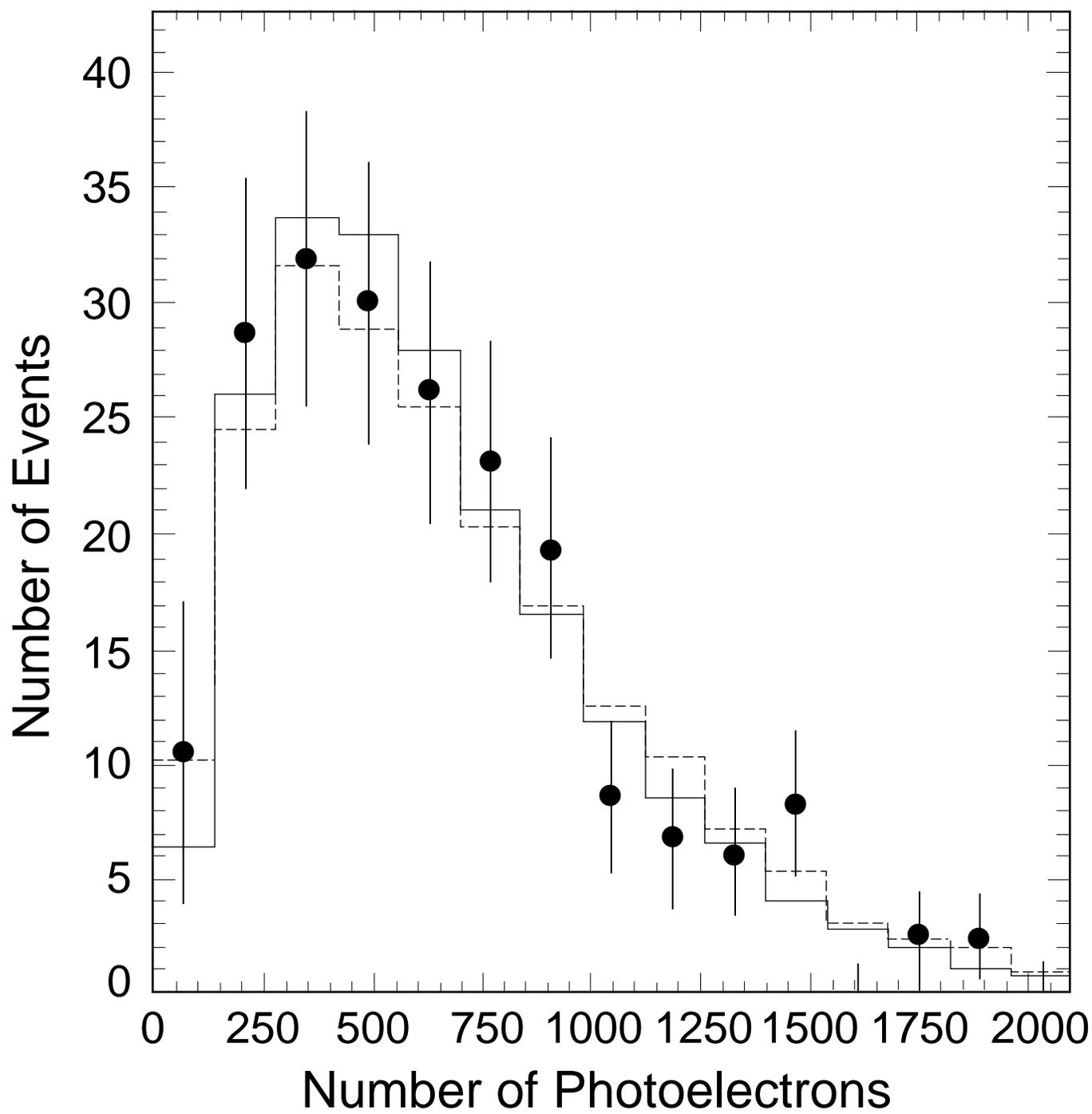
\end{document}